\documentclass[sigconf,natbib=true]{acmart}
\usepackage{multirow,multicol}
\usepackage{enumitem}
\usepackage{booktabs,url,tcolorbox,mathtools} 
\usepackage[stable]{footmisc}
\usepackage{newtxtext,newtxmath}
\usepackage{caption}
\acmConference{Unpublished preprint}{March}{2024}
\acmBooktitle{Proceedings of xxx}
\copyrightyear{2024}
\acmYear{2024}
\setcopyright{acmlicensed}
\acmPrice{XX.XX}
\acmDOI{XX.XXXX/XXXXXXX.XXXXXXX}
\acmISBN{XXX-X-XXXX-XXXX-X/XX/XX}
\renewcommand\footnotetextcopyrightpermission[1]{} 

\begin{document}
\title{A Comparison of Methods for Evaluating Generative IR}

\author{Negar Arabzadeh}
\affiliation{%
  \institution{University of Waterloo}
  \country{Canada}}

\author{Charles L. A. Clarke}
\affiliation{%
  \institution{University of Waterloo}
  \country{Canada}}

\begin{abstract}

Information retrieval systems increasingly incorporate generative components. For example, in  a retrieval augmented generation (RAG) system, a retrieval component might provide a source of ground truth, while a generative component summarizes and augments its responses. In other systems, a large language model (LLM) might directly generate responses without consulting a retrieval component. While there are multiple definitions of generative information retrieval (Gen-IR) systems, in this paper we focus on those systems where the system's response is not drawn from a fixed collection of documents or passages. The response to a query may be entirely new~---~text never seen before that may never be seen again. Since traditional IR evaluation methods break down under this model, we explore various methods that extend traditional offline evaluation approaches to the Gen-IR context. Some of these methods have been previously explored in a generative context, while others have been proposed but have not been developed and validated in a consistent framework. Offline IR evaluation traditionally employs paid human assessors, but increasingly LLMs are replacing human assessment, demonstrating capabilities similar or superior to crowdsourced labels.  Given that Gen-IR systems do not generate responses from a fixed set, we assume that methods for Gen-IR evaluation must largely depend on LLM-generated labels. Along with methods based on binary and graded relevance, we explore methods based on explicit subtopics, pairwise preferences, and embeddings. We first validate these methods against human assessments on several TREC Deep Learning Track tasks; we then apply these methods to evaluate the output of several purely generative systems. For each method we consider both its ability to act autonomously~--~without the need for human labels or other input~---~and its ability to support human auditing. To trust these methods, we must be assured that their results align with human assessments. In order to do so, evaluation criteria must be transparent, so that outcomes can be audited by human assessors. We provide all our code, prompts, and data for our experiments at \url{https://github.com/Narabzad/genir-evaluation}.

\end{abstract}

\maketitle

\section{Introduction}

Over the past two years, Generative Information Retrieval (Gen-IR) has emerged as a distinct paradigm for IR systems~\cite{min-etal-2023-factscore,sigir2023genir,rethinking,bohnet2023attributed,webglm,semnani-etal-2023-wikichat,jiang-etal-2023-active,shi2023replug,yu2023generate,gienapp2023evaluating}.
While there are many definitions for Gen-IR, 
we focus on those system where the system's response to a query is generated and not drawn from a fixed corpus. The scope of our work includes Retrieval Augmented Generation (RAG) systems, in which a retrieval component provides a source of ground truth, while a generative component based on a large language model (LLM) summarizes and augments retrieved results. Our scope also includes models that directly answer questions and queries, without depending on an external source of ground truth. These models include services like ChatGPT when it is responding to the type of informational query that would traditionally be directed to a web search engine. We exclude from our scope those generative systems where the output of the generative component is a document identifier. While we also consider these system to be Gen-IR systems, the response seen by the user is no different than that of a traditional search engine, apart perhaps from its quality~\cite{yang-etal-2023-auto,10.1145/3583780.3614821,pradeep-etal-2023-generative,sun2023learning,10.1145/3539618.3591631}. In this paper, we focus on those generative systems that fill the role of traditional search engines for informational queries, but whose response to a query may be entirely new~---~text never seen before that may never be seen again.

\begin{table*}[ht]
\centering
\begin{tabular}{|l|l|l|l|l|l|}
\hline
\textbf{Method}     & \textbf{Description}      & \textbf{Measure}   & \textbf{R1 - Agreement}   & \textbf{R2 - Auditable?} & \textbf{R3 - Autonomous?}\\ 
\hline \hline
Binary relevance    & Section \ref{meth:bin}    & Binary             & Figure \ref{fig:valid-binary}   & Yes                      & Yes \\
Graded relevance    & Section \ref{meth:grade}  & Ordinal scale      & Figure \ref{fig:valid-graded} & Uncalibrated             & Yes \\
Subtopic relevance  & Section \ref{meth:sub}    & Binary by subtopic & Figure \ref{fig:valid-sub}   & Yes                      & Yes \\
Pairwise preference & Section \ref{meth:pair}   & Binary             & Figure \ref{fig:valid-pref}  & Yes                      & Requires exemplar \\
Embeddings          & Section \ref{meth:embed}  & Cosine similarly   & Figure \ref{fig:valid-embed} & Indirect                 & Requires exemplar \\
\hline
\end{tabular}
\caption{Summary of methods compared. }
\label{tab:methods}
\end{table*}

Under the traditional information retrieval paradigm, a search engine responses to a query with a ranked list of items drawn from a corpus. While the number of possible responses is large, these possible responses are known in advance. In particular, the performance of the search engine's ranking component can be measured in terms of its ability to place more relevant items higher in the ranked list. Traditional offline evaluation of search engines employs paid human labellers to examine query/page pairs and rate them according to their relevance, quality, and other factors~\cite{ars08,judges08,ly11,voor98}. A key benefit of offline assessment is its repeatability. Once sufficient query/page pairs are labeled, new ranking models can be compared to previous models, and the long-term performance of a search engine can be tracked over weeks and months.

This form of evaluation remains dominant in academic information retrieval research and experimental forums, including TREC, NTCIR, CLEF and FIRE. One common goal of these forums is to create reusable test collections, consisting of a corpus of items, a set of queries, and a set of relevance labels for item/query pairs~\cite{10.1145/3269206.3271766,voorhees2022old,craswell2021trec}. When a research team creates a new ranking model, its performance can then be directly compared to the performance of historical models. One unfortunate aspect of this approach occurs when a new model returns an unjudged result \cite{arabzadeh2024fr}. The typical default is to assume it is not relevant, potentially leading us to underestimate the performance of the new model. In ideal circumstances, these assessment ``holes'' would immediately be filled by additional human labels \cite{arabzadeh2022shallow}. While this ``hole filling'' may be possible for large commercial search engines \cite{arabzadeh2023adele}, in academic contexts the expense and need for consistency with existing labels usually precludes it \cite{arabzadeh2023quantifying}.

During the same two-year period that saw the emergence of Gen-IR, LLMs have begun to augment and replace human assessors for offline evaluation~\cite{Gilardi_2023, thomas2023large, Faggioli2023LLM,holefilling}. Most notably, \citet{thomas2023large} report on their experience with Bing, writing, ``We have found large language models can be effective, with accuracy as good as human labellers... we find that models produce better labels than third-party workers, for a fraction of the cost, and these labels let us train notably better rankers.''
LLM-based assessment can also successfully fill holes, potentially allowing traditional test collections to better reflect performance improvements~\cite{holefilling}.
If we accept that LLMs are an adequate, or even superior, replacement for human assessors, we create an apparent circularity in which LLM-based Gen-IR systems are primarily evaluated by LLM-based assessors. Considering this situation in early 2023, \citet{Faggioli2023LLM} write, ``LLMs are not people. IR measures of effectiveness are ultimately grounded in a human user’s relevance judgment... In a plain old search engine, results for a query are ranked according to predicted relevance... Each has a clear source, and each can be inspected directly as an entity separate from the search engine. '' Any proposal to employ LLMs for evaluating Gen-IR systems must retain this grounding in human judgment, even if LLMs become the primary source of relevance assessments.

In this paper, we explore multiple methods for evaluating Gen-IR systems that could serve as a replacement for the traditional IR test collections. Since the corpus is no longer fixed, the test collection would consist solely of queries and additional relevance criteria specific to that query. To evaluate a new model, the query would be issued to the Gen-IR system and its response would be evaluated according to the relevance criteria. While this evaluation process is entirely ``hole filling'', new responses can become part of the relevance criteria. For example, we might retain a ``best known response'' to be compared with future responses. Given the success of LLM-based labeling, we assume that labeling will be LLM based, without the need for human input, but that human assessors will audit LLM-generated labels, so that the process continues to be grounded in human judgment. Since the labels will be audited, relevance criteria should be clearly defined, so that human assessments and LLM-based labels remain consistent.
This assumption reflects the process described by \citet{thomas2023large}, who do not require human input, but do ``retain human oversight and audit examples of LLM output, although we do not audit every label.''

In exploring methods for evaluating Gen-IR systems, we consider the following requirements:
\begin{itemize}
\item[R1.]
\textit{Agreement with human assessment:} When applied to existing test collections, the methods should produces outcomes that are consistent with human assessors. 
\item[R2.]
\textit{Suitability for auditing by human assessors:} Relevance criteria should be sufficiently defined so as to allow a human assessor to easily confirm that an LLM has correctly applied the relevance criteria to the response.
\item[R3.]
\textit{The ability to operate autonomously:} Apart from auditing, the method should not require input from human assessors, including an explicit explanations of what is, or is not, relevant, or even exemplars of ideal responses~\cite{bohnet2023attributed}.
\end{itemize}
R1 embodies the most basic requirement of any IR evaluation method, i.e., that it should align with human expectations. R2 provides the mechanism by which we ensure that R1 holds.
R3 eliminates the need for human input beyond the audit requirement of R2. Some of the methods we consider require one or more ideal responses as exemplars. Others might benefit from explicit statements of what a response should contain, or not contain, in order to be considered relevant. If human input is required to add a query to our test collection it creates a bottleneck, causing delays and limiting the number of queries we can add. 

One important requirement we do not consider is the correctness of the response. LLMs are widely known to ``halluncinate'', producing convincing but factually incorrect output. This problem has been explored by others, including efforts to check the output of LLMs against knowledge bases and other sources of ground truth~\cite{min-etal-2023-factscore, feng-etal-2023-factkb,liu2023evaluating,Huo_2023}. In a deployed evaluation system, checking the factual accuracy of the response would be an independent step, performed by a separate evaluation component. In this work, we are focused on the relevance of the response. Separating correctness from relevance helps to ensure that all information provided in the response~---~even information that is incidental to relevance~---~is factually correct.

Our work provides an experimentally focused complement to the recent theoretical analysis provided by \citet{gienapp2023evaluating}. Like us, they focus on the contrast between traditional evaluation over a finite corpora vs. evaluation over the ``infinite index'' provided by a Gen-IR system. They view Gen-IR systems as a fourth generation of search engine that directly supports synthetical search tasks, in which material from multiple sources is compiled and synthesized into a new document that directly answers a query. Although the scope of their theoretical analysis extents beyond the scope of our methods and experiments, we take the initial steps towards operationalizing their vision of Gen-IR evaluation.

\begin{figure*}[t]
\begin{tcolorbox}[colback=gray!5!white,colframe=gray!75!black]
You are a helpful TREC assessor. 
\newline
Provide the subtopics for a given query in the same format.  \newline
Example 1:  ... \newline
\newline
Example 2:  \newline
<topic number="254" type="faceted"> \newline
<query>barrett's esophagus</query> \newline
<subtopic number="1"> Find the causes and risk factors of Barrett's Esophagus. </subtopic> \newline
<subtopic number="2"> What treatments are available for Barrett's Esophagus? </subtopic> \newline
<subtopic number="3"> What is the recommended diet for Barrett's Esophagus? </subtopic> \newline
<subtopic number="4"> Is there a link between Barrett's Esophagus and cancer? </subtopic> \newline
<subtopic number="5" > Find symptoms of Barrett's Esophagus. </subtopic> \newline
<subtopic number="6"> Find pictures of Barrett's Esophagus. </subtopic> \newline
<subtopic number="7"> Find Barret's Esophagus studies performed at the Mayo Clinic. </subtopic> 
\newline
</topic> \newline
 \newline
Example 3:
...
  \newline
\newline
  Question: \{query\}
  \newline
  Passage: \{passage\}
\end{tcolorbox}
\caption{Prompt for generating subtopics. Examples are elided for space.}
\label{fig:answer_prompt}
\end{figure*}

\section{Methods}
\label{sec:methods}

Table~\ref{tab:methods} lists the methods compared in this paper. In this section we describe each method, along with any associated background and related work.  In Section~\ref{sec:valid} we validate each method in terms of its agreement with human assessment (R1). In addition, for each method, we consider both the suitability for auditing (R2) and the ability to act autonomously, without the need for example responses, ideal responses or explicitly defined relevance criteria (R3).

\subsection{Binary relevance}
\label{meth:bin}

This method is perhaps the simplest possible. We prompt an LLM to compare a query/response pair, asking if the response is relevant\footnote{We have made all our prompts publicly available at \url{https://github.com/Narabzad/genir-evaluation}}. Auditing by human assessors is straightforward, since binary relevance is already standard for IR evaluation, going back to the earliest days of the Cranfield experiments~\cite{10.1145/122860.122861}. For auditing (R2) we ask the human assessor the same question that we ask the LLM. Under this method, there is no need for additional relevance criteria beyond the query itself. If one or more ideal responses are available they may be provided as examples for few-shot prompting, but they are not required (R3).

We make a distinction between this method and the evaluation methods employed for some question answering collections, where system responses must match a predefined, factual answer, perhaps expressed in several ways~\cite{bulian-etal-2022-tomayto, ouyang2022training}.
Traditionally, those collections required either an exact match between the predefined ``gold'' answer or responses are measured with F1.
More recently, LLMs have been applied to match responses against the gold answers~\cite{bulian-etal-2022-tomayto, ouyang2022training}. In this paper, we assume that a query requires more than just a straightforward, factual answer. It might be answered in many different ways, with different responses containing different information, so that we measure binary relevance, rather than a match to a gold answer.


\subsection{Graded relevance}
\label{meth:grade}

This method is the graded equivalent of the previous method. We prompt an LLM to consider a query/response pair and assign it one of several relevance grades. The prompt will include definitions for each grade. For auditing (R2) we again ask the human assessor the same question that we ask the LLM. If example responses for each grade are available, they may be included for few-shot prompting, but they are not required (R3). Both \citet{thomas2023large} and \citet{Faggioli2023LLM} provide examples of prompts, with \citet{thomas2023large} also providing comparative experiments over a range of prompts. While the prompt in \citet{Faggioli2023LLM} includes examples, the prompts in \citet{thomas2023large} do not.

One concern about use of graded relevance by both human and LLM assessors is the need for calibration on the definitions of the various grades.  When looking at a single query/response pair in isolation, it may be unclear what grade should be assigned to the response. In general, it is difficult to precisely define the distinction between ``relevant'', ``highly relevant'' and ``perfect'' results, for example, and it may not be possible to create definitions that work for all queries~\cite{mmst17}. Even for binary relevance, agreement between human assessors can be low~\cite{voor98}, especially when assessors did not originate the query and are not experts on its topic~\cite{judges08,kyct13}. While binary relevance may also need calibration as to what should and should not be considered relevant, the use of graded levels increases this need.


\subsection{Subtopic relevance}
\label{meth:sub}

We use the term query ``subtopics'' to mean \textit{a set of predicates that can be applied to responses, such that more relevant responses are expected to satisfy more subtopics.} For example, given the query ``What are the applications of robotics in the world today?'', \citet{10.1145/860435.860440} consider each possible application of robotics to be a separate subtopic (e.g., ``spot-welding robotics'' or ``controlling inventory - storage devices''). In other prior research, the term ``nuggets'' is used to mean what we consider to be an equivalent concept. Given a query on the assassination of JFK, \citet{nuggets} suggest, ``John Kennedy was elected president in 1960.'' as a possible nugget. Older TREC Web Tracks use the term ``faceted query'' for the same concept. Given a query about the singer Neil Young, \citet{trecweb} suggest ``Find albums by Neil Young to buy,'' as a possible facet. To make an assessment in terms of a subtopic (or nugget or facet) the assessor determines if a document ``contains'' (information about) it, a binary label.

To develop subtopics, we prompt an LLM to create them (Figure~\ref{fig:answer_prompt}). For the experiments reported in this paper, our prompt includes several examples based on queries from older TREC Web Tracks~\cite{trecweb}. To assess relevance we prompt the LLM to make a binary assessment for each subtopic (plus the query itself). To assign a relevance level to a response, we simply count the number positive judgments and normalize by the number of subtopics (plus one for the query itself)~\cite{10.1145/860435.860440}. Like the previous methods, we can ask the LLMs and human auditors the same question for each subtopic (R2). Unlike graded relevance, subtopic assessments require only binary judgments, reducing the need for calibration, while providing more than one level of relevance. It is also possible for a human assessor to audit the subtopics themselves, adding or removing subtopics to provide a more precise and complete measure of relevance, but this step is entirely optional (R3). 

\citet{nuggets} describe and validate a proposal for nugget-based evaluation that aligns with the method in this section. In the initial ``Construction Phase'' of their proposal, assessors would identify nuggets in relevant documents, expressing them as sentences, fragments or patterns. Together the queries and the nuggets represent a reusable test collection, with relevance defined by the nuggets. In the ``Evaluation Phase'' of their proposal, nuggets are matched against document contents. For matching purposes \citet{nuggets} experiment with shingle matching, although they suggest other possible approaches.

While \citet{nuggets} build a convincing case for nugget-based evaluation, the proposal was not widely adopted. The proposal required training for human assessors to extract nuggets, and the matching process was limited to exact matches at the shingle level. With LLMs now available to define nuggets and match them against documents at more than just the lexical level, we may now be able fully realize the potential of this proposal to simplify the creation of reusable collections.

\subsection{Pairwise preferences}
\label{meth:pair}

Pairwise preferences provide a simple way to capture human feedback that ensures better inter-assessor agreement~\cite{kyct13,cvs21,cbcd08,sz20,xie20}. Human preference labels are also central to the reinforcement learning from human feedback (RLHF) process used to tune LLMs and other models~\cite{christiano2023deep}. For evaluation, we prompt an LLM to determine which of two responses is best. For auditing, we ask the human assessor the same question (R2). However, to assess a response, we need something to compare it against. For each query in our test collection we need at least one exemplar response for comparison purpose. While previous system responses could provide exemplars, this approach provides no way of knowing when a Gen-IR system is failing completely. We may be comparing one bad response against another, creating nothing but noise.

Due to the need for an exemplar response we hesitate to characterize this method as fully autonomous (R3). For each query, we might employ a human assessor to write an exemplar response, or we might retrieve potential responses from a corpus and ask a human assessor to judge their relevance. We also might employ one of the methods above (binary, graded or subtopic relevance) until an adequately relevant exemplar is found, and then switch to pairwise preferences. For the experiments reported in this paper, we use relevant passages from an existing test collection as exemplars. 

\subsection{Embeddings}
\label{meth:embed}

Recently, \citet{arabzadeh2024adapting} propose and validate an evaluation method based on the cosine similarity between the embeddings of an exemplar and a generated response, where more relevant responses are assumed to have greater similarity with the exemplar. This approach differs from the preceding approaches in that we are not prompting an LLM to make a judgment or comparison, but rather using it to generate embeddings. That work was inspired by BERTScore~\cite{zhang2020bertscore}, which takes a similar approach to evaluating summarization and translation. Like pairwise preferences, the need for an exemplar causes us to hesitate in characterizing this method as fully autonomous (R3). In the strictest sense, auditing is not possible (R2), since we cannot expect a human assessor to generate an embedding. However, indirect auditing is possible through consistency with any of the other methods, particularly pairwise preference judgments.

\section{Validation}
\label{sec:valid}
As discussed in Section~\ref{sec:methods} Gen-IR evaluation methods based on binary relevance, graded relevance, and embeddings have been well validated in the research literature~\cite{thomas2023large, arabzadeh2024adapting,holefilling}. In this section we provide additional validation for these methods, as well as providing validation for methods based on subtopics and preference labels, which have received less attention for Gen-IR evaluation. Data, prompts and code for our validation experiments are released at~{\url{https://github.com/Narabzad/genir-evaluation}}.

\subsection{Datasets}

Our validation experiments are conducted with passages from the TREC 2019 and TREC 2020 Deep Learning Track datasets (DL 2019 and DL 2020)~\cite{DL19,DL20}\footnote{\url{https://microsoft.github.io/msmarco/TREC-Deep-Learning.html}}. The DL 2019 dateset comprises 43~queries and 9260~relevance judgments (``qrels''). The DL 2020 dataset comprises 54 queries and 11386~qrels. Passages are judged on a four-point scale: 0 (``Irrelevant''), 1 (``Related''), 2 (``Highly relevant'') and 3 (``Perfectly relevant''). For both datasets, the corpus comprises 8.8 million passages derived from the MS MARCO V1 corpus~\cite{10.1145/3404835.3462804}\footnote{\url{https://microsoft.github.io/msmarco/}}.

\subsection{Models}
\label{valid:models}

To assess binary relevance, graded relevance, subtopics, and preferences we used OpenAI's {\tt gpt-4} and {\tt gpt-3.5-turbo} models accessed though its public API in January 2024. Apart from the partial prompt in Figure~\ref{fig:answer_prompt}, we do not include prompts in this paper for reasons of space, but they are included in our publicly available GitHub repository. For embeddings (Section~\ref{valid:embed}) we used Vanilla BERT embeddings employed for the experiments reported in \citet{arabzadeh2024adapting}. The experiments reported in that section build on the experiments of \citet{arabzadeh2024adapting} with additional analysis.

\subsection{Methodology}

Traditional IR evaluation focuses on measuring effectiveness on a ranked list~\cite{kk02}. Effectiveness measures, such as NDCG, map relevance labels into gain values to compute an aggregate effectiveness score over items in the ranked list. Average effectiveness scores are then computed over query sets.

Since Gen-IR systems do not operate over fixed collections, they generally do not produce ranked lists. Instead, they might generate only a single response. While some current systems (e.g., ChatGPT) may generate alternative responses that the user can scroll between, they do not generate long ranked lists. Since a Gen-IR system may produce only a single response, it becomes important to measure the effectiveness of this response. If we are comparing two or more models to each other, we should be able to determine if the performance of one model on a particular query is better or worst than that of another. 
As as result, we focus our validation on the extent to which each of the methods in Section~\ref{sec:methods} is able to recognize performance differences between responses. We ask:
\begin{quote}
    \textit{If human assessment indicates that one system's response to a query is better than another system's response, how often does the evaluation method agree with that assessment?}
\end{quote}
The specific relevance grade, subtopic count, or cosine similarity is not important because we are not computing an aggregate score over a ranked list.
To help recognize performance differences, for each query in DL~2019 and DL~2020 we group the qrels (i.e., passage/judgment pairs)
into three categories:
\begin{itemize}[labelwidth =\widthof{{\em Unacceptable:}}, leftmargin = !]
\item[{\em Best known:}]
For each query, these are the qrels at the top grade for that query. Some queries may have no qrels at grade 3, or even grade 2, although all queries have at least one relevant qrel.
\item[{\em Acceptable:}]
For each query, these are the relevant qrels that fall between the grade of the best known qrels and the judged non-relevant qrels (grade~0).
\item[{\em Unacceptable:}]
For each query, these are the judged non-relevant qrels (grade~0).
\end{itemize}
Each method is measured in terms of its ability to make assessments that are consistent with these categories. It is particularly important that methods distinguish the best known results and other results. If we expect the performance of Gen-IR systems to exceed that of traditional search engines, recognizing differences between relevant results becomes more important than recognizing non-relevant results, since we expect these systems to rarely generate non-relevant responses.

\begin{figure}[p]
\centering
  \includegraphics[clip, trim=6.1cm 2.5cm 1cm 0.466cm,scale=0.51]{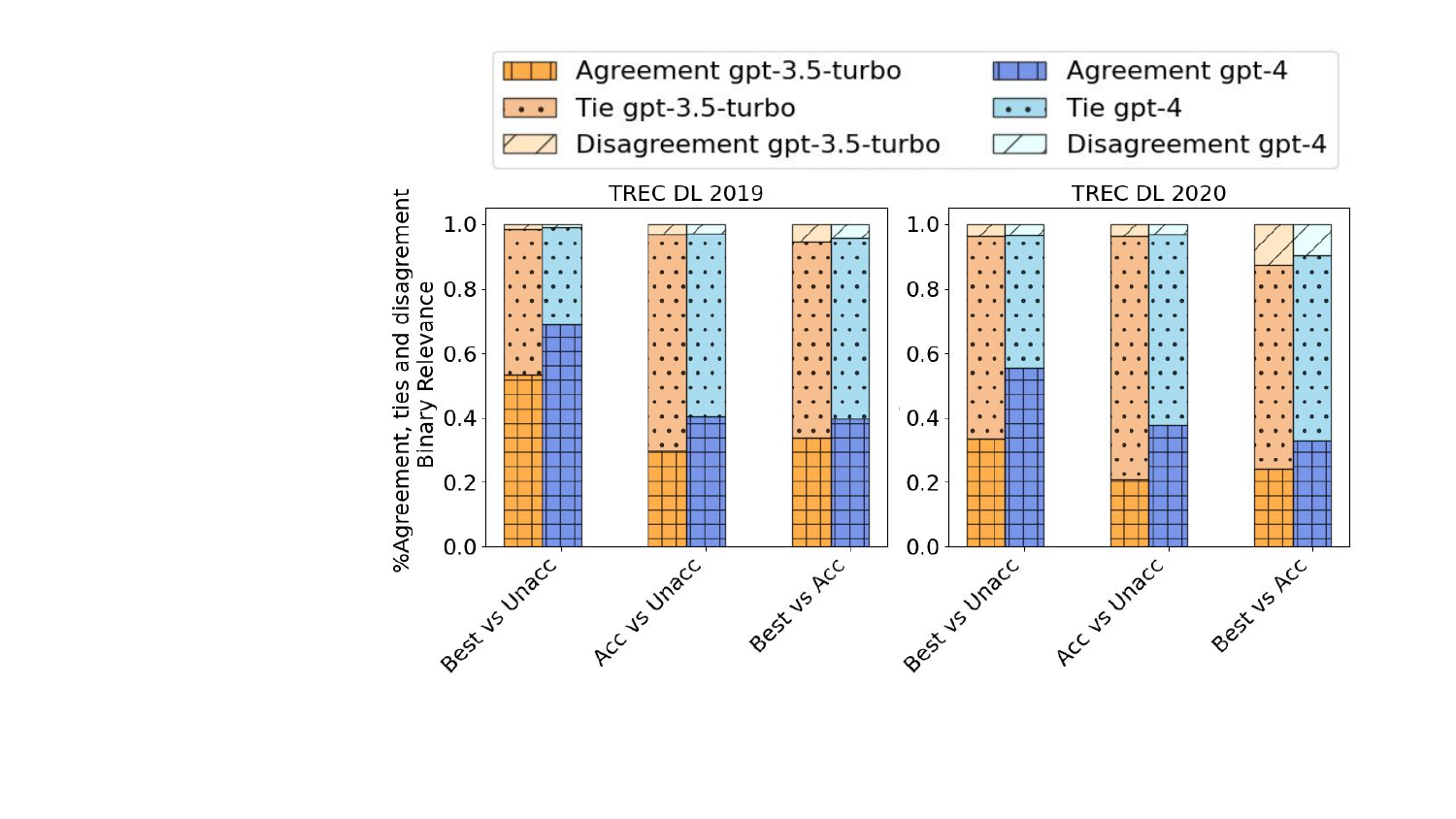}
\caption{Validation of binary relevance.}
\label{fig:valid-binary}
\end{figure}

\begin{figure}[p]
\centering
  \includegraphics[clip, trim=6.1cm 2.5cm 1cm 0.466cm,scale=0.51]{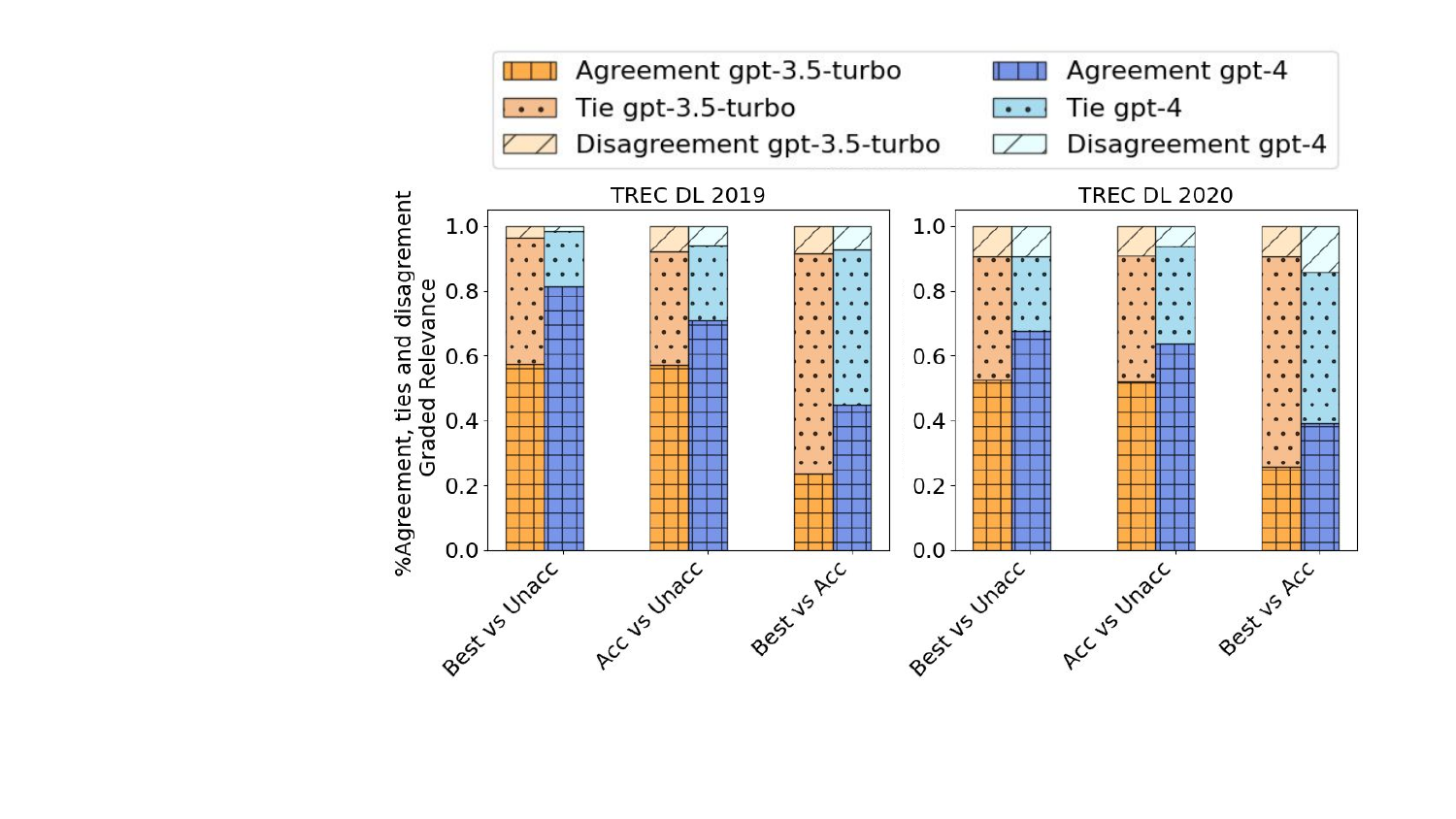}
\caption{Validation of graded relevance.}
\label{fig:valid-graded}
\end{figure}

\begin{figure}[p]
\centering
  \includegraphics[clip, trim=6.7cm 2.5cm 1cm 0.466cm,scale=0.53]{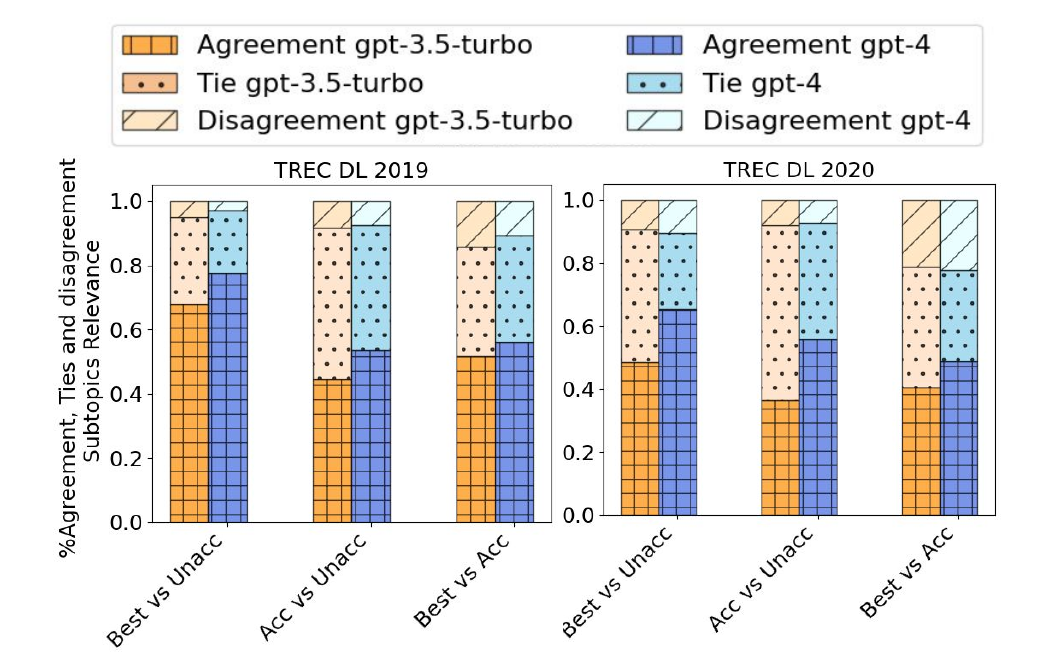}
\caption{Validation of subtopic relevance.}
\label{fig:valid-sub}
\end{figure}

\subsection{Results}
\label{valid:results}\label{valid:bin}\label{valid:grade}\label{valid:sub}\label{valid:pair}\label{valid:embed}

Figure~\ref{fig:valid-binary} plots validation results for binary relevance (Section~\ref{meth:bin}). For any given query, for each pair of qrels from different categories we compare binary relevance, as determined by LLM assessment, and check for agreement with the ordering implied by the catgories Best (\texttt{Best}), Acceptable (\texttt{Acc}) and Unacceptable (\texttt{Unacc}). For example, if one qrel is in the \texttt{Best}  category and one qrel is in the \texttt{Unacceptable} category, the pair is in agreement if the \texttt{Best} qrel is relevant and \texttt{Unacceptable} qrel is not. If they are both relevant or both non-relevant they tie. Otherwise, they are in disagreement.


Figure~~\ref{fig:valid-binary} shows relative few disagreements, but many ties. If binary relevance indicates that one system is better than another, we might reasonably trust that outcome, but if the results are tied we do not know. Figure~\ref{fig:valid-graded} plots validation results for graded relevance. When compared to binary relevance, graded relevance produces higher agreement, fewer ties, but slightly greater disagreement. In both cases, {\tt gpt-4} produces fewer ties than {\tt gpt-3.5-turbo}, consistent with the expectation that {\tt gpt-4} provided better performance than {\tt gpt-3.5-turbo} on zero-shot tasks.

\begin{figure}[t]
\begin{minipage}{\linewidth}
\noindent {\bf Query 1105792}: define: geon
\begin{itemize}
\item What is the definition of a geon?
\item What is the origin of the term geon?
\item What are examples of geons in cognitive psychology?
\item How are geons used in object recognition?
\item What are the limitations or criticisms of the geon theory?
\end{itemize}
\vspace*{\baselineskip}
\noindent {\bf Query 1108651}: what the best way to get clothes white
\begin{itemize}
\item What are the best laundry detergents for white clothes?
\item What are some home remedies to whiten clothes?
\item How to use bleach safely for white clothes?
\item What are the washing machine settings for white clothes?
\item How to remove stains from white clothes?
\item What are the best fabric conditioners for white clothes?
\item How to maintain the whiteness of clothes over time?
\end{itemize}
\vspace*{\baselineskip}
\noindent{\bf Query 1109707}: what medium do radio waves travel through
\begin{itemize}
\item How do radio waves travel through space?
\item Can radio waves travel through water?
\item How do radio waves travel through the atmosphere?
\item What materials can block or interfere with radio waves?
\item How does the medium affect the speed and distance of radio wave travel?
\end{itemize}
\end{minipage}
\caption{Examples of LLM-generated subtopics.}
\label{fig:subtopic-eg}
\end{figure}

\begin{table}[t]
\centering
\begin{tabular}{|l|l|c|c|c|c|}
\hline
\textbf{Track} & \textbf{Model}         & \textbf{Min} & \textbf{Max} & \textbf{Mean} & \textbf{Median} \\ \hline \hline
DL 2019          & \texttt{gpt-4}                  & 5            & 8            & 5.72          & 5               \\ 
DL 2019          & \texttt{gpt-3.5-turbo}          & 2            & 7            & 5.21          & 5               \\ 
DL 2020          & \texttt{gpt-4}                  & 4            & 8            & 5.69          & 5               \\ 
DL 2020          & \texttt{gpt-3.5-turbo}          & 1            & 8            & 4.94          & 5               \\ \hline
\end{tabular}
\caption{Statics for the number of LLM-generated subtopics.}
\label{tab:subtopic-stat}
\end{table}

Figure~\ref{fig:valid-sub} plots validation results for subtopic relevance. In some cases, the number of ties and disagreements are slightly higher with graded relevance; in other cases, slightly lower.
Figure~\ref{fig:subtopic-eg} shows several examples of subtopics generated by the prompt in Figure~\ref{fig:answer_prompt}. Statistics for the number of LLM-generated subtopics, on a per-query basis, appears in Table~\ref{tab:subtopic-stat}.  All generated subtopics are included in our data release.
As we see in these examples, the subtopics often include partial answers or relevant information not appearing in the query. The subtopics for Query~1105792 (``define: geon'') reflect the meaning of this term in the context of Irving Biederman's recognition-by-components theory from cognitive psychology. The subtopics for Query~1108651 (``what the best way to get clothes white'') suggest laundry detergent and bleach as possible solutions. The subtopics for Query~1109707 (``what medium do radio waves travel through'') include space, water, and the atmosphere as possible media. The processes for nugget based evaluation defined by \cite{nuggets} proposes that human assessors extract nuggets from relevant documents. At least in the case of these relatively straightforward and factually oriented queries, the LLM was able to identify aspects of relevance without consulting relevant documents. Furthermore, this assessment strategy offers greater explainability and interpretability as it can highlight specific subtopics in which the document either successfully addresses or falls short of addressing.

\begin{figure}[t]
\centering
  \includegraphics[clip, trim=7cm 2.5cm 1cm 0.9cm,scale=0.51]{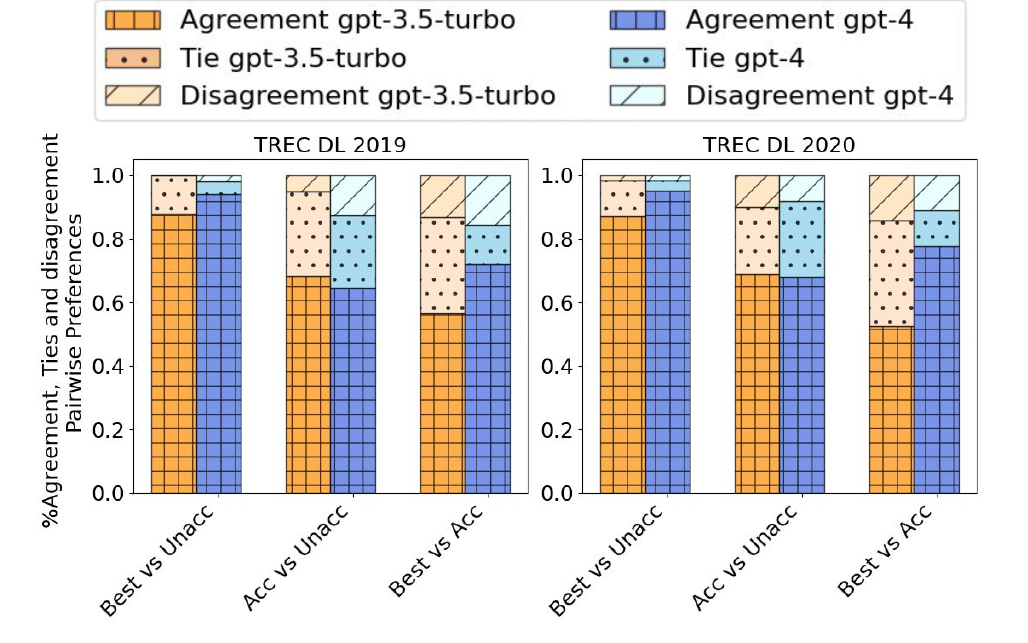}
\caption{Validation of pairwise preferences.}
\label{fig:valid-pref}
\end{figure}

\begin{figure}[t]
\centering
  \includegraphics[clip, trim=1.9cm 1.5cm 3cm 1cm,scale=0.42]{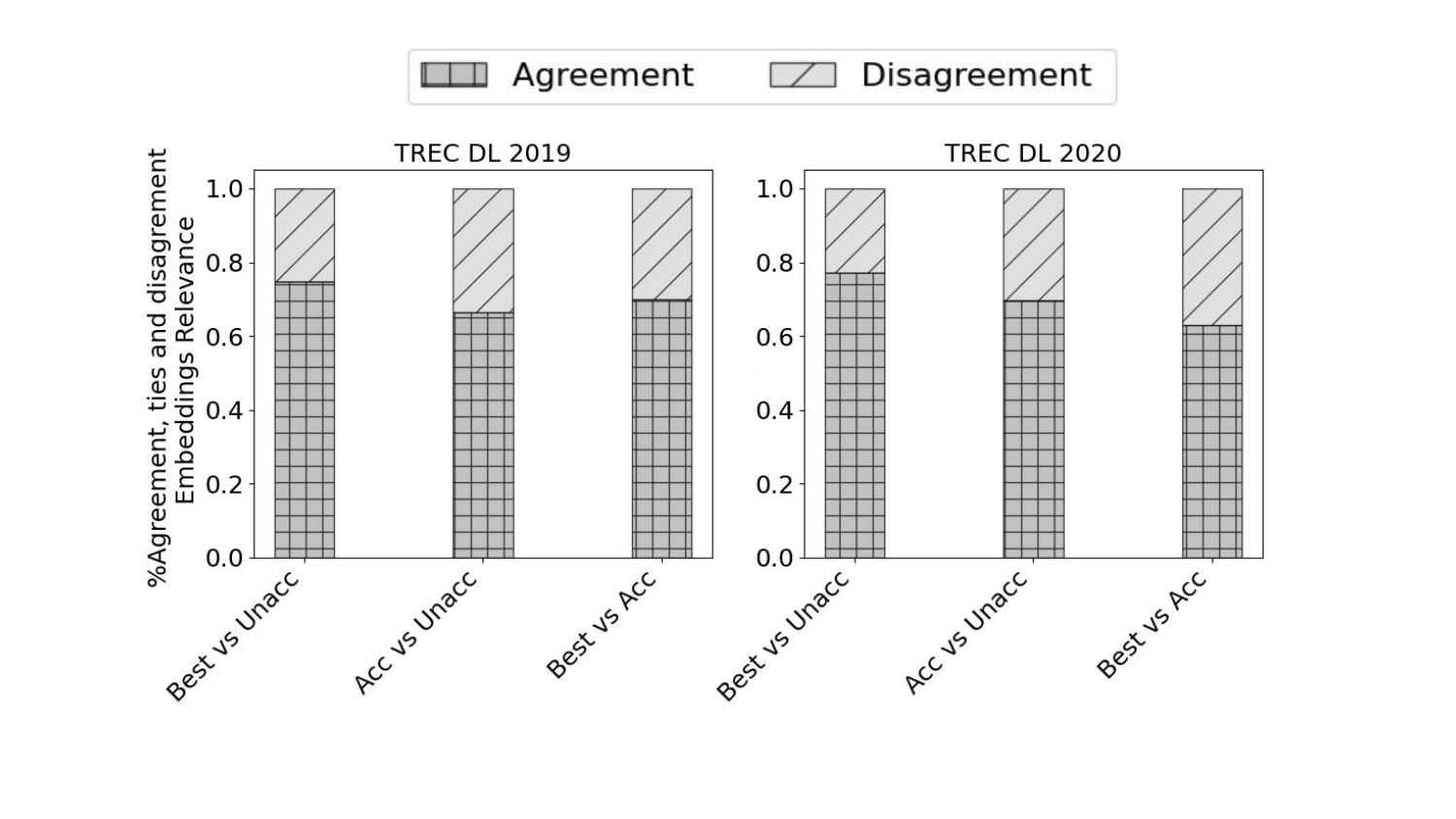}
\caption{Validation of embeddings with Vanilla BERT.}
\label{fig:valid-embed}
\end{figure}

To validate pairwise preferences, we first create sets containing query/qrel/qrel triples of the form $(\textit{query id}, \textit{category}_1, \textit{category}_2)$, where $\textit{category}_1$ and $\textit{category}_2$ are distinct relevance categories from $\{\texttt{Best}, \texttt{Acceptable}, \texttt{Unacceptable}\}$. Categories are ordered so that no pair of qrels appears in more than one set of triples. Since this process creates millions of triples we randomly sample five triples from each set. For each sampled triple, we prompt a LLM to determine which qrel is more relevant to the query. Since LLMs are known to be sensitive to the order of results in a pairwise prompt~\cite{qin2023large}, we prompt twice with the qrels in different orders. If the preferred qrel changes with the ordering, we treat it as a tie. Figure~\ref{fig:valid-pref} plots validation results for pairwise preferences. When compared to other methods (including embeddings Figure~\ref{fig:valid-embed}) there are fewer disagreements and ties. Since a pairwise preference directly compares one result to another, instead of operating through an intermediating relevance grade or similarity, this outcome may not be surprising, but it confirms the potential of pairwise preferences as an evaluation method.

Validation of embeddings is based on the same triples. For each triple of query and two distinct qrels from two different categories $(\textit{query id}, \textit{category}_1, \textit{category}_2)$, we pick a third qrel from the \texttt{Best} category that does not appear in the triple. We treat this third qrel as an exemplar and compute similarity between it and the embeddings of the qrels in the triple. The qrel with the greatest similarity is the preferred qrel. For embeddings similar to \citet{arabzadeh2024adapting}, we used the Vanilla BERT embeddings. Since embeddings for all qrels are available and could be compute offline, computing cosine similarity is fast, and unlike pairwise preferences, we do not samples triples. Figure~\ref{fig:valid-embed} plots validation results for embeddings. Due to the nature of cosine similarity, ties rarely occur with embeddings, and do not occur at all in this experiment. Results are superior to binary, graded, and subtopic relevance, but overall, they are inferior to preferences, especially \texttt{gpt-4} perferences.

\section{Gen-IR Experiments}
\label{sec:gen} \label{sec:expr}

\begin{figure}[th]
\centering
  \includegraphics[clip, trim=0cm 1.3cm 7.7cm 0.1cm, scale=0.5]{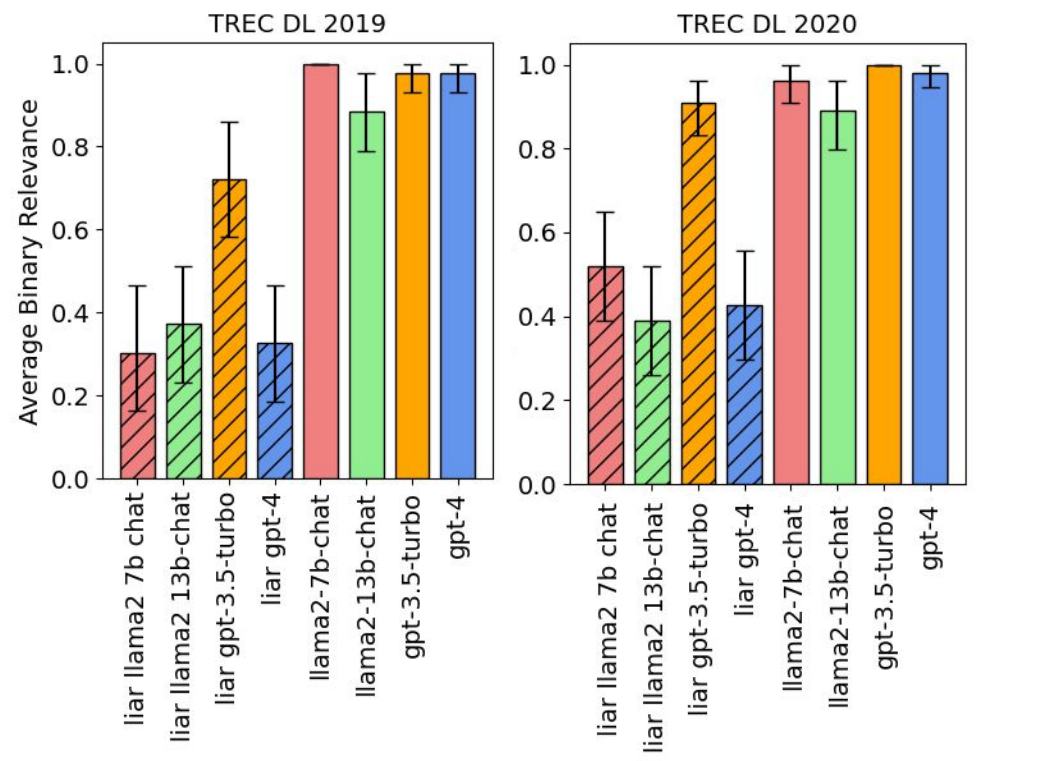}
  \caption{Evaluation of generative responses based on binary relevance. Each bar represents the average performance with a 95\% confidence interval based on 1000 bootstrap samples. The confidence interval on some bars is too small to be visible.}
  \label{fig:gen-accuracy}

  \includegraphics[clip, trim=0cm 1.2cm 6cm 0cm, scale=0.5]{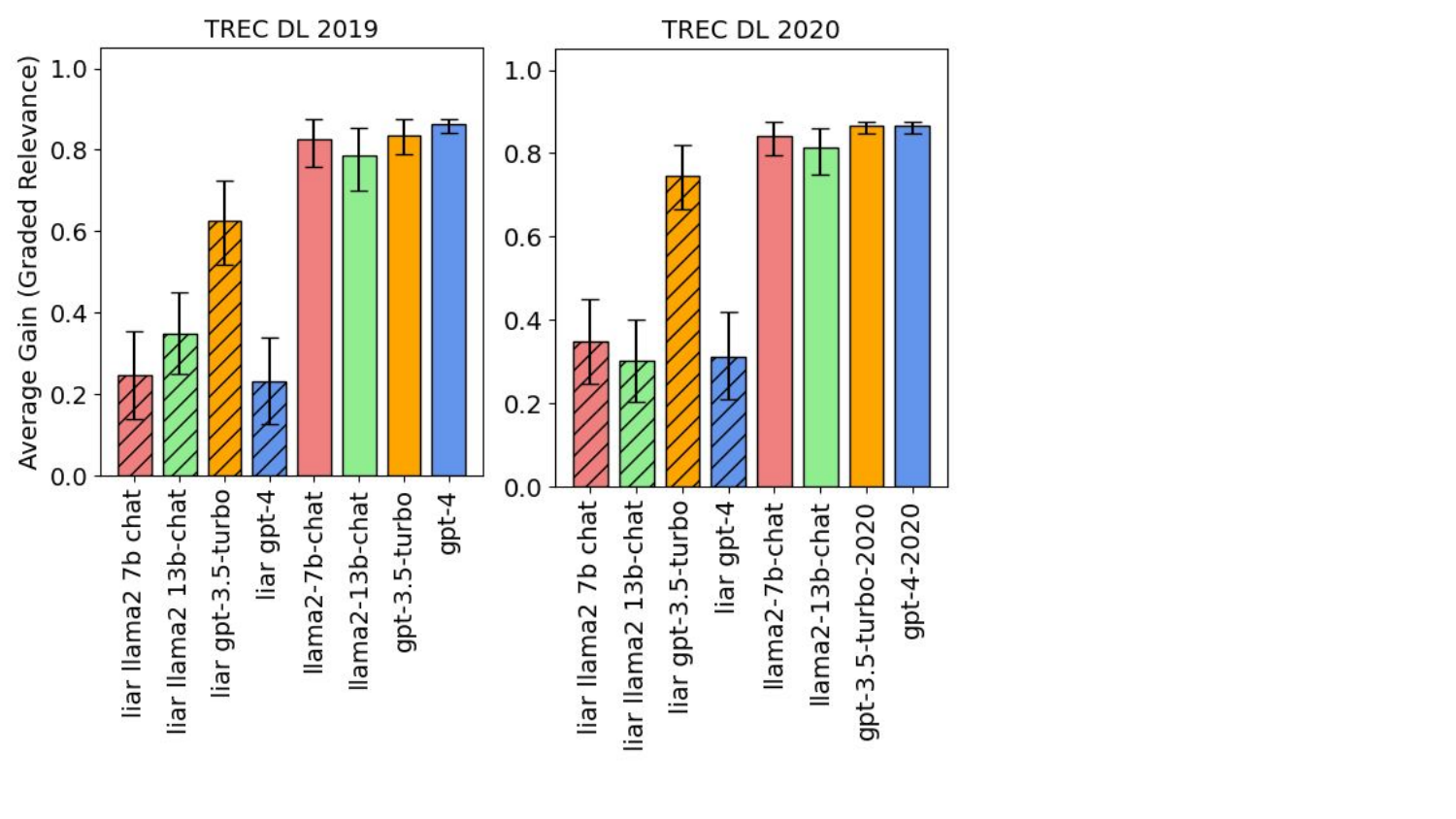}
  \caption{Evaluation of generative responses based on graded relevance.}
  \label{fig:gen-graded}
\vspace{-1em}
\end{figure}

In this section, we explore the efficacy of the five different evaluation methods detailed in Section~\ref{sec:methods} and validated in Section~\ref{valid:results}. To achieve this, we assess the generated responses by four distinct language models, which we treat as Gen-IR systems~\cite{arabzadeh2024adapting}. These models include {\tt gpt-3.5-turbo} and {\tt gpt-4}, representing commercial LLMs with lower and higher numbers of parameters, respectively. Additionally, we employ {\tt llama2} to investigate the capabilities of open-source LLMs for this task, using both a smaller variant ({\tt llama2 7b chat}) and a larger one ({\tt llama2 13b chat}), as described by \citet{touvron2023llama}. Having a larger and smaller version of the same model is beneficial for our experiments since it is commonly assumed in various downstream applications in NLP and computer vision that a larger model is more likely to provide superior responses compared to a smaller one\cite{pavlichenko2022best,roberts2020much}.

To provide a contrast to these four LLMs, we follow a similar approach to \citet{arabzadeh2024adapting} by creating \textit{liar} versions, where the LLM is prompted to provide incorrect answers to questions. The prompts for this scenario, as well as all the other prompts, can be found in our github repository. This technique involves deliberately prompting LLMs to produce incorrect responses, and it provides a couple of advantages:   Firstly, it allows us to conduct an additional validation for the evaluation methods, as we expect these models to demonstrate significantly lower performance compared to models prompted to generate correct answers. Secondly, it provides insights into the models' creative abilities when generating responses that may appear correct but are ultimately incorrect. In the experimental results that follow, the four liar versions of the LLMS are called {\tt liar llama2 7b chat}, {\tt liar llama2 13b chat}, {\tt liar gpt-3.5-turbo} and {\tt liar gpt-4}. \citet{arabzadeh2024adapting} provides some examples of responses from these liars, which can be entertaining.

In the experiments, we will assess the average performance of all eight LLMs treated as Gen-IR systems  responding to the queries in DL 2019 and 2020 according to the five different evaluation methods defined and valiadated in previous sections. Since the experiments in Section~\ref{sec:valid} show that using a larger LLM ({\tt gpt-4} vs {\tt gpt-3.5-turbo}) provides greater agreement with human preferences, we adopt {\tt gpt-4} for all the evaluation related experiments in this section.
We note that for all the experiments, we report the average performance on all the queries in each of the TREC DL 2019 and 2020 dataset with 95\% confidence intervals using bootstrap sampling with 1000 samples. We note that the confidence interval on some of the one bar plots might be too small to be visible.

\subsection{Binary Relevance}

In Figure \ref{fig:gen-accuracy}, we plot the performance of generated responses under binary relevance. We present the percentage of generated responses that were judged relevant by the LLM assessor. As seen in the figure, for the TREC DL 2019 dataset, all LLMs significantly outperformed their respective "liar" versions.  
Additionally, we observe that binary relevance may not be sufficient to distinguish between larger and smaller models. On both datasets both {\tt gpt-3.5-turbo} and {\tt gpt-4} are observed to have similar performance. When it comes to {\tt llama}-based models, the larger model (13b) is seen to have lower performance than the smaller one (7b). 

\subsection{Graded Relevance}

In Figure \ref{fig:gen-graded}, we plot the performance of generated responses under graded relevance. We convert relevance grades to gains values using the standard exponential NDCG formulation.  When comparing Figure \ref{fig:gen-accuracy} with Figure \ref{fig:gen-graded}, we observe that the confidence intervals for graded relevance are sightly narrower than those for binary relevance. However, in terms of the ability to distinguish between models, the overall findings are consistent.

\subsection{Subtopics}

\begin{figure}
  \includegraphics[clip, trim=0cm 1cm 7cm 0cm,scale=0.5]{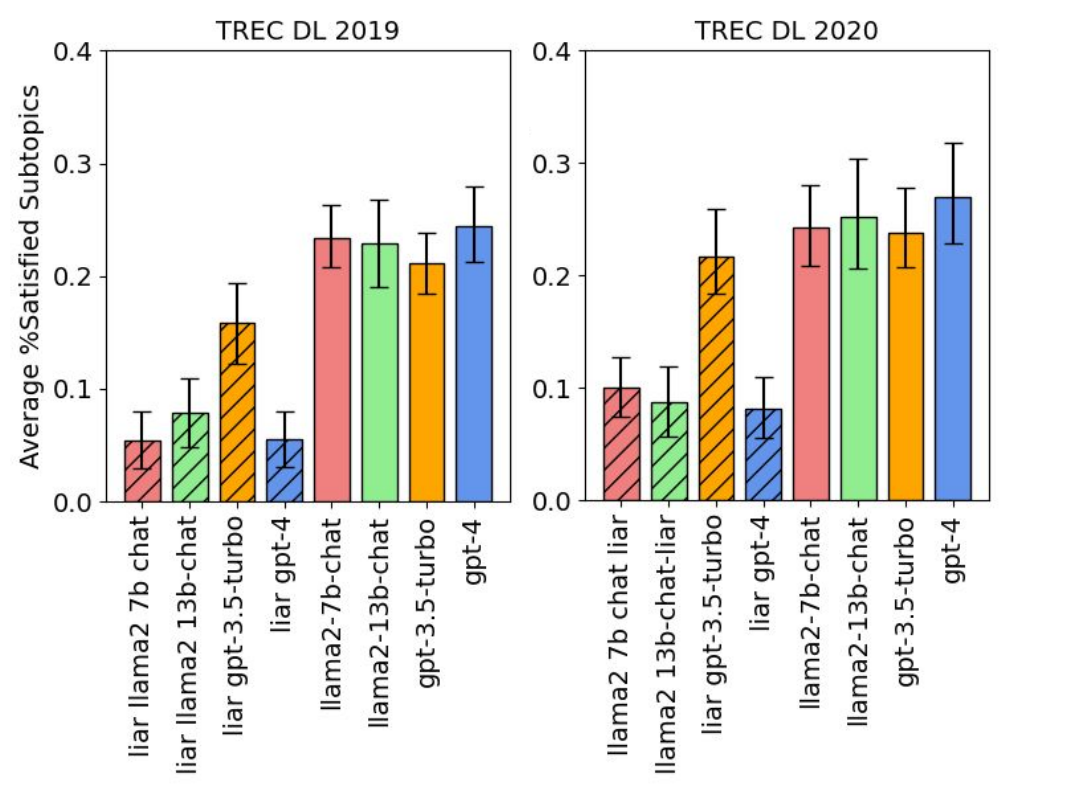}
\caption{Evaluation of generative models based on subtopics.
}
  \label{fig:gen-subtopics}
\end{figure}

In Figure~\ref{fig:gen-subtopics}, we plot the average percentage of relevant subtopics for each generative model. The subtopics for each query are derived from Section \ref{valid:sub} using the {\tt gpt-4} model for each dataset. Unlike graded relevance and binary relevance assessments, which yield a limited number of discrete values, utilizing the percentage of answered subtopics as the evaluation metric provides a larger set of values. Consequently, we observe a wider confidence intervals in Figure~\ref{fig:gen-subtopics} when compared to Figure~\ref{fig:gen-accuracy} and Figure~\ref{fig:gen-graded}.
Additionally, {\tt gpt-4} exhibits superior average performance compared to {\tt gpt-3.5-turbo}. This difference in performance between the two {\tt gpt} models is not observed in either binary or graded relevance assessments.

\subsection{Pairwise Preferences}

As indicated in Table~\ref{tab:methods}, the assessment of Gen-IR with pairwise preferences and embeddings necessitates the use of at least one exemplar per query. For evaluating the responses generated by LLMs using the pairwise preferences approach, we employ the following procedure: For each query, we randomly select one of the qrels from the \texttt{Best} category, which might be level 2 (``Highly relevant'') or level 3 (``Perfectly relevant'') depending on the query. Subsequently, we determine the percentage of comparisons where the generated answer is preferred over the exemplar, which serves as the best-known answer to date. 

To eliminate bias in our comparisons, we repeat each pair of comparisons twice. First by presenting the LLM-generated response followed by the exemplar, asking the LLM which answer they prefer. Then we perform the same comparison with the LLM-generated answer and exemplar positions swapped to ensure there is no positional bias in the responses.
Similar to the experiments described in Section~\ref{valid:pair}, when both trials agree, we consider that answer as the preferred one. However, if the preference changes when flipping the LLM-generated answer and exemplar, we categorize that example as a tie, and it does not count as the LLM's answer being preferred over the exemplar. Figure~\ref{fig:gen-pref} plots the percentage of queries where the generate responses was preferred over the exemplar. This process could potentially be repeated with all the qrels in the \texttt{Best} category and the average percentage of times the generated responses were preferred over the top tier answer could be reported. However, due to the experimental cost, we only performed this assessment with a randomly selected qrel.

As shown in Figure~\ref{fig:gen-pref}, pairwise preferences are effective in distinguishing between regular LLMs and their "liar" versions. In fact, the difference between regular and liar promptings is most pronounced with this assessment method. Furthermore, the disparity between {\tt gpt-4} and {\tt gpt-3.5-turbo} is more pronounced, and is statistically significant. While the pairwise preference method offers performance benefits, it is relatively costly due to the large number of pairwise comparisons that may be required. As an additional observation, on TREC DL 2020 the response of {\tt gpt-4} was almost always preferred over the exemplar. In a production environment where ongoing assessment is taking place, perhaps on a daily basis, these responses might replace or augment the exemplar.

\subsection{Embeddings}

\begin{figure}[t]
\centering
  \includegraphics[clip, trim=0cm 1.8cm 9.5cm 0cm,scale=0.53]{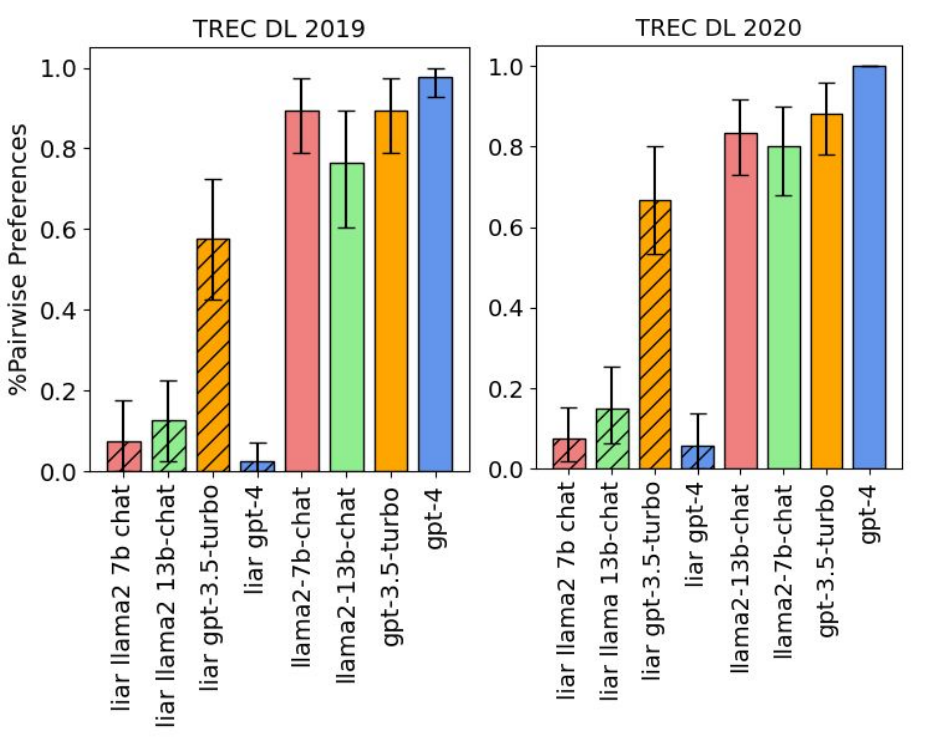}
\caption{Evaluation of generative models based on pairwise preferences. Each bar represents the average performance with a 95\% confidence interval based on 1000 bootstrap samples. The confidence interval on one bar is too small to be visible.}
  \label{fig:gen-pref}

  \includegraphics[clip, trim=0cm 1.8cm 9cm 0cm,scale=0.52]{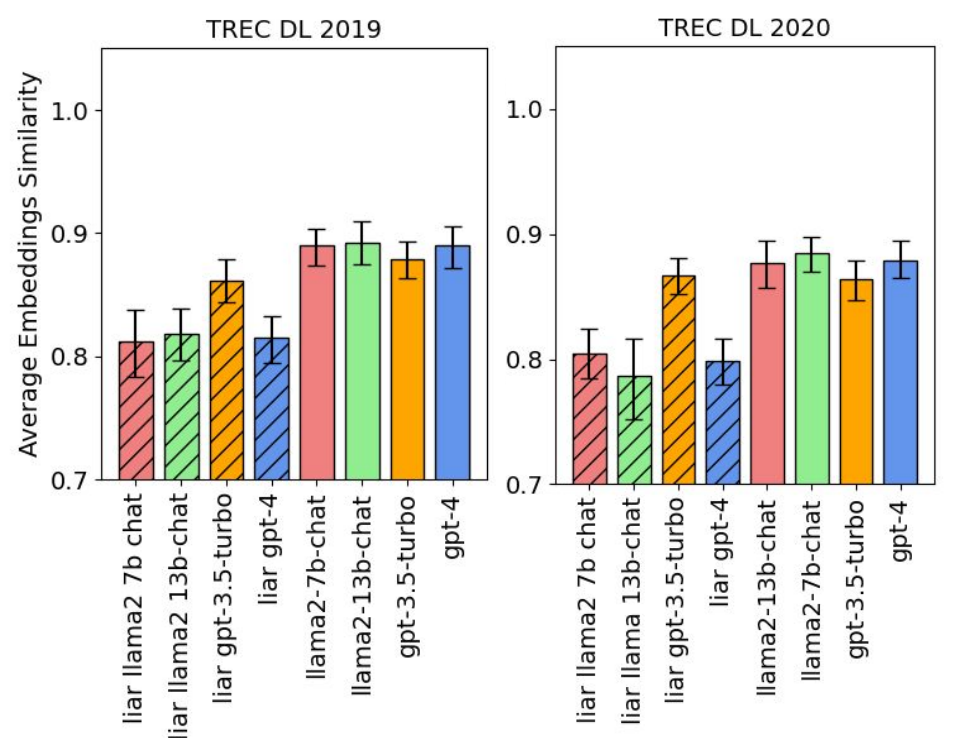}
\caption{Evaluation of generative models based on the embeddings.
}
  \label{fig:gen-emb}
\vspace{-2em}
\end{figure}

Similar to Pairwise preferences, using embedded representations of the generated responses for assessment requires an exemplar. To accomplish this, we calculate the average similarity of the generated responses to the best-known answers or qrels in \texttt{Best} category. For each query, we compute the average cosine similarity between the generated response and all the qrels in the \texttt{Best}  category. In Figure \ref{fig:gen-emb}, we plot the average of these similarities across all the queries in each dataset.
As expected, all models outperform their liar versions, but as seen in all plots in this section, \texttt{liar-gpt-3.5-turbo} outperforms all other liar models. Perhaps \texttt{gpt-3.5-turbo} is not good at lying, or perhaps we are not good at prompting it to lie.

\section{Limitations of this Work}

We recognize a number of important limitations in our work. First, our experiments are limited to two years of a single TREC Track, Deep Learning 2019 and 2020. In addition, we focus on the passage task of these tracks, and we do not consider full documents. As seen in Figure~\ref{fig:subtopic-eg}, many of the queries can be satisfied by relatively short, factual responses and do not require the synthetical search engine envisioned by \citet{gienapp2023evaluating}. The appearance of partial answers in the subtopics of Figure~\ref{fig:subtopic-eg} may reflect this property of the questions.

Like \citet{gienapp2023evaluating}, our work is limited to the traditional ``\textit{ad hoc}'' scenario, where we evaluate a single response to a query in isolation. In reality, we expect future Gen-IR systems to be conversational in nature, with most queries taking place in the context of a larger interaction~\cite{zamani2023conversational}. It is likely that responses will be personalized to the individual user~\cite{personal}. While we expect the methods explored in this paper to remain appropriate in conversational context, verifying that hypothesis would require further experimentation.

Like any work that prompts LLMs to perform zero-shot tasks our results depend on the exact prompts employed. Different prompts might produce different outcomes~\cite{thomas2023large}. Since we intend our evaluation methods to be auditable by human assessors (R2) we use naturalistic prompts, similar to the instructions we would give to a human. These prompts are provided in our data release.In addition, we only investigated two families of LLMs: ``{\tt gpt}'' as the representative of closed-source LLMs and ``{\tt llama2}'' as the representative of open-sourced LLMs. Further investigation with other LLMs is required to strengthen the generalizability of our findings.

Commercial LLMs accessed though APIs provide no way to precisely specify or control the model. The model called ``{\tt gpt-4}'' may have changed several times during short time span of days during which we conducted our experiments. The model named ``{\tt gpt-4}'' may even be multiple models fine-tuned in different ways, with the model changing from call to call according to A/B testing criteria. Our account may even be locked to an A/B testing bucket in which our ``{\tt gpt-4}'' performs better, worse, or just differently than the production model. By the time this paper appears at the conference, ``{\tt gpt-4}'' may be an entirely different model, or it may be deprecated and no longer accessible. In addition, the circularity of using ``{\tt gpt-4}'' as the assessor of ``{\tt gpt-4}'' might bring biases to our assessment.

\section{Conclusion}

Since Gen-IR systems do not generate results from a fixed collection, reusable test collections for Gen-IR cannot easily depend on human assessment.  As a result, we assume that LLMs will take the place of humans for primary assessment of Gen-IR system responses. Ideally they would act autonomously, with no input from humans regarding what is and is not relevant, but we also assume that humans will retain an auditing role for quality assurance.  Under these assumptions, we validate and experimentally explore five methods for evaluating Gen-IR systems: binary relevance, graded relevance, subtopic relevance, pairwise preferences, and cosine similarity between embeddings. Of these methods, subtopic relevance appears to provide a reasonable compromise between autonomy and auditabilty, with the LLM itself defining subtopics and the human auditor required only to make binary relevance decisions regarding individual subtopics. If an exemplar is available for comparison, pairwise preferences provide the best overall performance, both in our validation experiments and in recognizing differences between generative models. However, this approach is quite computationally expensive compared to the rest.

As immediate future work, we plan to continue our exploration over additional test collections and additional Gen-IR systems, especially RAG systems.
In addition, we plan to conduct a human study to validate our findings in section \ref{sec:valid} on the generated text i.e., Section \ref{sec:gen}. Extensive studies on human annotated data is required to investigate to what extend our findings apply to generated texts. 

Our ongoing work combines and extends the methods in this paper.
Pairwise preferences require exemplars, and we plan to explore an approach that combines subtopic relevance with pairwise preferences, where subtopic relevance first identifies potential exemplars and pairwise preferences selects the best of them. We imagine a future Gen-IR test collection consisting of a query set, subtopics for each query, and one or more exemplars of relevance. For RAG evaluation, the test collection might even contain a corpus as a source of ground truth for both result generation and result factchecking. In the longer term, we plan to extend this work to consider personalization, diversity, and other factors that are difficult to address under traditional IR evaluation approaches.

\newpage
\balance
\bibliographystyle{ACM-Reference-Format}
\bibliography{references}

\end{document}